# Mechanical Softening of Vero Cells Induced by an Attenuated Measles Vaccine Virus

Alexander Einschütz López[1#], Johanna Bacher[2#], Luis N. Ponce-Gonzalez[1], and José L. Toca-Herrera*[1]

[1]Institut für Biophysik, NWNR, Universität für Bodenkultur Wien, Austria.

[2] Institute of Bioprocess Science and Engineering, Department of Biotechnology, BOKU University, Vienna, Austria

#Both authors contributed equally to this work

*Corresponding author: Jose L. Toca-Herrera, e-mail: jose.toca-herrera@boku.ac.at

## Graphical abstract

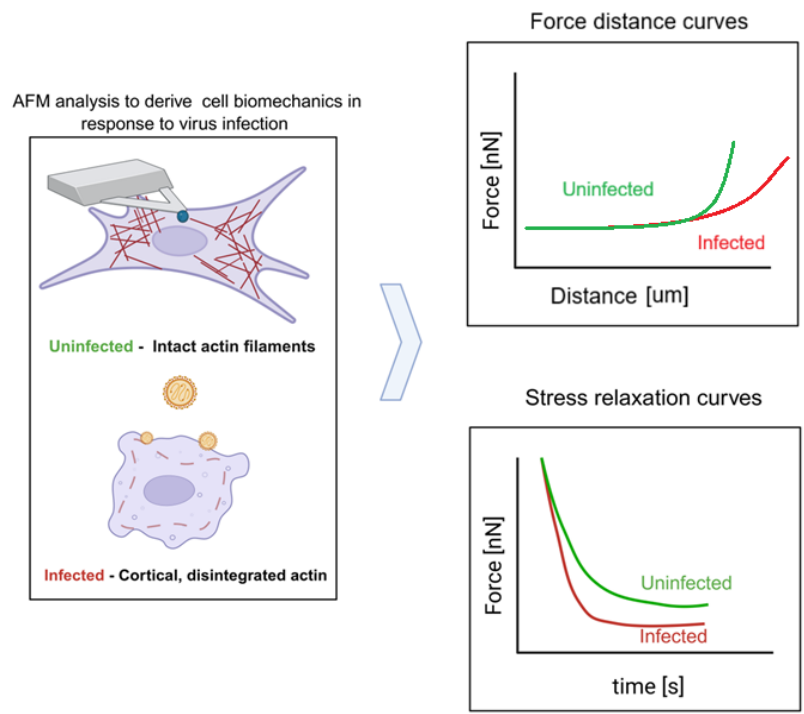

## Abstract

Quantitative characterization of biophysical alterations caused by viral infection remains at an early stage. In this study, we examined the mechanical response of Vero cells after exposure to an attenuated Measles Vaccine Virus (MVV) using atomic force microscopy (AFM) in combination with confocal microscopy. AFM force–distance measurements were conducted in the perinuclear region to evaluate changes in elastic and viscoelastic properties. Within 24 hours post-infection (hpi), cells infected at a multiplicity of infection (MOI) of 0.5 exhibited an approximately 35% decrease in median Young's modulus relative to uninfected controls, indicating substantial cellular softening. Corresponding shifts in viscoelastic behavior were observed, including reductions in the relaxed modulus $E_0$ and in viscosities $\eta_1$ and $\eta_2$ (effects comparable to those induced by cytochalasin-D–mediated actin depolymerization). Confocal microscopy further revealed an MOI-dependent reorganization of the cytoskeleton, marked by altered F-actin distribution and changes in filament architecture. Together, these findings suggest that actin remodeling contributes to the altered viscoelastic properties observed during infection. Overall, this work proposes a straightforward and complementary approach for characterizing virus–cell interactions by integrating AFM with confocal imaging.

## Keywords

## 1. Introduction

Viral infection induces pronounced alterations in cellular morphology and mechanics, affecting parameters such as cortical stiffness, membrane tension, and cytoskeletal organization (1, 2). Consequently, developing methods to characterize this biophysical remodeling is essential for identifying early stages of infection, cell damage, or viral release.

The cytoskeleton is central to maintaining cellular integrity and mechanical stability. Actin filaments, microtubules, and intermediate filaments collectively govern cell shape, intracellular transport, and structural resilience. Tensegrity-based models have shown that cytoskeletal prestress and force balance regulate the mechanical behavior of cells (3). Among cytoskeletal components, the cortical actin network is the primary determinant of the cell's elastic response (4, 5). Notably, rearrangements of the cytoskeletal architecture occur during viral entry, replication, and egress (6–8). Despite this, quantitative biophysical data describing infection-induced shifts in mechanical properties—such as stiffness, viscoelasticity, and stress-relaxation behavior—remain limited. This gap is significant given that mechanical parameters often change sensitively and early, sometimes preceding visible morphological alterations associated with cellular stress or pathology (9–11).

Atomic force microscopy (AFM) has emerged as a powerful method for probing single-cell mechanics (12). AFM force–distance measurements enable precise quantification of the apparent Young's modulus (13), while force–time measurements allow extraction of viscoelastic properties, including relaxation times and viscosities (14, 15).

In this study, an attenuated Measles Vaccine Virus (MVV) was used as an infection model system (16-20). Measles virus (MV) belongs to the Paramyxoviridae family, an enveloped, negative sense single-stranded RNA virus, known to trigger characteristic innate immune responses and cause cytopathic effects (16, 21). Infection experiments were carried out using Vero CCL81, which is a frequently used cell line for viral research and production (22). The efficiency of virus propagation in these cells is not only influenced by biochemical factors (23), but also by their mechanical properties such as Young's modulus (24).

Despite the substantial knowledge about the biological effects of the MVV infection, the mechanical responses of Vero cells remain largely uncharacterized. To address this, the present work combines AFM with confocal microscopy to quantify changes in Vero cells 24 hours after infection with MVV. By measuring both elastic properties and viscoelastic response and correlating with cytoskeletal architecture, the aim was to identify a biophysical signature of viral infection.

## 2. Materials and methods

The content of this section can be found in the Section S1 of the Appendix.

## 3. Results

### 3.1. Infection with MVV induces softening of Vero cells

The mechanical properties of Vero cells were measured using AFM. Single cells were identified via bright-field imaging, and force-distance curves were recorded above the nuclear region. The indentation and the Young modulus were calculated with Equation S1 and Equation S2, respectively, after considering the effect of the substrate (see Appendix, Section S.1.4.1).

Our findings demonstrate that exposure and infection with the model virus at an MOI of 0.5 led to a significant decrease in cellular stiffness 24 hours post-infection (hpi). Figure 1A shows that the mean Young's modulus (E) decreased to 421 ± 13 Pa, compared to 702 ± 19 Pa in uninfected control cells (mock). This mechanical response to infection was comparable to that in cells treated with 5 µM cytochalasin D (i.e., E=321 ± 17 Pa), an actin depolymerization agent, which was used as a positive control. Infection at a lower viral dose (MOI 0.1) did not result in a statistically significant change compared to not-infected cells at 24 hpi, with a mean Young's modulus of 643 ± 16 Pa. In agreement with the inverse relationship between stiffness and deformation, indentation depths showed a corresponding trend toward greater deformation for the treated groups (Figure 1B).

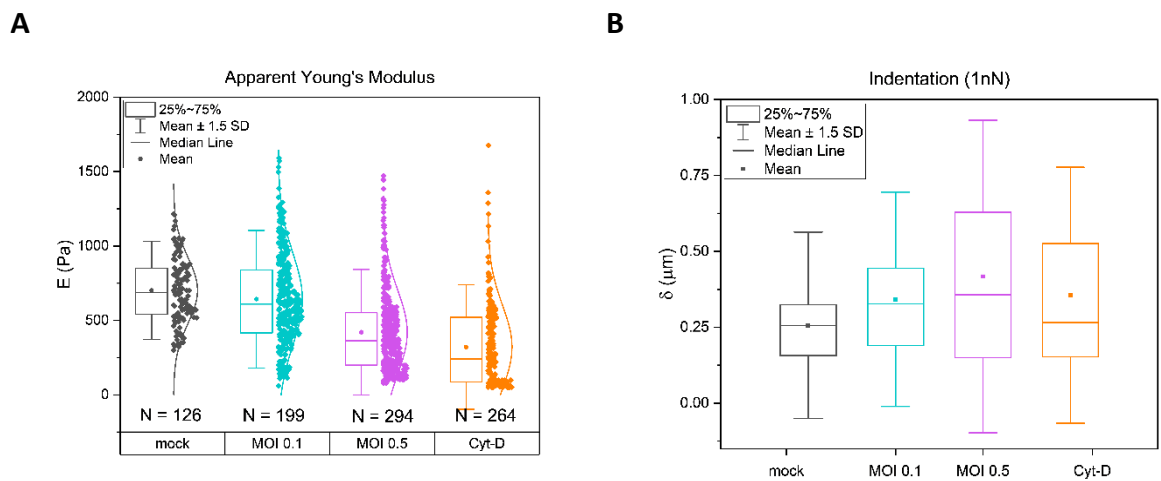

**Figure 1**. Exposure of Vero cells to MVV. Influence on the apparent Young's modulus, E (A), and the indentation (δ) at 1 nN force (B) on measurements performed in Vero cells. An indentation depth of 500 nm was used to calculate the Young's modulus. Box plots show the interquartile range (first to third quartile), with whiskers representing 1.5× the standard deviation. The central point indicates the mean, and the black line indicates the median. MOI 0.5 and Cyt-D groups are significantly different from MOI 0.1 and mock ($p < 0.001$). The indentations of all tested treatments are significantly different compared to mock for $p < 0.05$.

### 3.2. Viscoelastic properties after virus infection from stress relaxation experiments

Force-time curves at a constant height were measured to obtain viscoelastic properties from stress-relaxation responses. The constant height pause segment (force relaxation) was fitted to a 5-element Maxwell model (Equation S6), as well as to a phenomenological power law model (Equation S8).

Fittings using the Maxwell model provided information about the elastic moduli ($E_i$), the equilibrium modulus ($E_\infty$) as well as the relaxation times ($\tau_i$) and viscosities ($\eta_i$). The fit yielded two elastic moduli and two distinct relaxation times, a short-term relaxation time of < 1 s and a long-term relaxation time of < 10 s (Table S1, see Appendix Section S2). Fittings using the power-law yielded two parameters: relaxed modulus ($E_0$), and power-law exponent ($\beta$).

Figure 2A shows that for the first viscous dashpot ($\eta_1$), mock exhibits the highest median viscosity (10.8 ± 0.9 Pa s). All treated conditions display lower median viscosity values (MOI 0.1: 6.3 ± 0.8; MOI 0.5: 6.8 ± 0.8; Cyt-D: 4.8 ± 0.4 Pa s). Differences between conditions are reflected primarily in shifts in the median values and changes in the interquartile ranges. A similar pattern is observed in the second viscous dashpot ($\eta_2$). Note that the scale is two orders of magnitude higher. Within this group, the control condition again shows the greatest spread, while the treatment conditions present narrower distributions and lower median viscosities.

Figure 2B shows that mock cells displayed the highest relaxed modulus ($E_0$ = 288 ± 10 Pa), which is consistent with a predominantly elastic response, as well as a low $\beta$ value (0.072 ± 0.019), which indicates limited viscous behavior. Cells infected at an MOI of 0.1 exhibited similar elastic properties to the uninfected cells with ($E_0$ = 237 ± 13 Pa and; $\beta$ = 0.081 ± 0.002). In contrast, cells infected at an MOI 0.5 and cells treated with Cyt-D presented a lower relaxed modulus ($E_0$ = 147 ± 7 Pa and $E_0$ = 129 ± 8 Pa; respectively), accompanied by a modest significant increase in fluidity ($\beta$ = 0.084 ± 0.002 and $\beta$ = 0.096 ± 0.002, respectively).

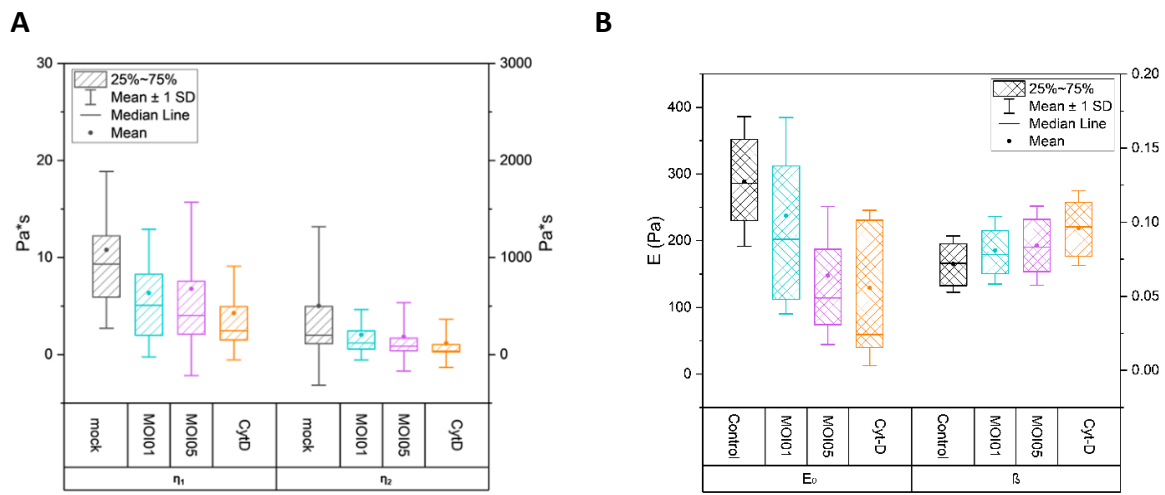

**Figure 2.** Exposure of Vero cells to MVV. Influence on the viscosities ($\eta_i$) extracted from fitting experimental data with Maxwell model (A) and on the relaxed modulus ($E_0$) and the exponent (ß) from power-law model (B). Box plots represent the data from the first to the third quantile with the whisks being 1x the standard deviation. Dot represents mean values and the black line represents the median. Viscosity values decrease systematically in the modified conditions relative to control, indicating reduced viscous dissipation and faster stress relaxation across the corresponding Maxwell elements. Treated groups are statistically significant compared to mock ($p < 0.02$). $E_0$ and $\beta$ values are statistically different from each other except in ß: mock – MOI 0.1. The exponent $\beta$ characterizes the viscoelastic relaxation spectrum: values approaching $\beta=0$ correspond to predominantly elastic, solid-like behavior, whereas values approaching $\beta=1$ indicate increasingly fluid-like behavior dominated by viscous dissipation.

### 3.3. *MVV infection leads to altered cell morphology and cytoskeleton*

We visualized effects of the model virus on the organization of the cytoskeleton and actin filaments under the tested conditions by using confocal fluorescence microscopy (Figure 3). In uninfected cells (Figure 3A), the Phalloiding staining showed a robust and highly organized network of filamentous actin, which was characterized by stress fibers spanning the cytoplasm and maintaining distinct cell morphology. This morphology was indicative of an intact cytoskeletal structure as well as mechanical integrity. In contrast, treatment with Cyt-D caused notable disruption to the cytoskeleton (Figure 3B), with fragmented or absent actin. This treatment also led to nonregular morphologies of the cells, which was consistent with the destabilization of the cortical actin framework.

Infected cells displayed dose-dependent cytoskeletal alterations. At a lower MOI of 0.1 (Figure 3C), actin accumulated at the cell periphery with partial loss of internal stress fibers, indicating cortical reorganization. Following, cells infected at a higher MOI (0.5) exhibited more extensive remodeling, characterized by a large absence of internal actin filaments and the formation of dense cortical actin rings along the cell periphery (Figure 3D). The cytoskeletal remodeling observed in infected cells at higher MOI was accompanied by increased cellular deformability, in agreement with AFM measurements.

Quantitative analysis of confocal images revealed a significant reduction in cell area at MOI 0.5 and in Cyt-D treatment, with mean areas of 864 ± 293 µm² and 692 ± 201 µm², respectively (Figure S1A). In contrast, mock and MOI 0.1 exhibited substantially larger cell areas (1628 ± 673 µm² and 1782 ± 860 µm², respectively). On the other hand, the mean fluorescence intensity (MFI), reflected the staining of actin filaments by Phalloidin per cell (Figure S1B). Mock showed the highest signals of MFI per cell, while Cyt-D treatment, MOI 0.1 and MOI 0.5 reduced the global fluorescence. Following, the maximal fluorescence intensity (Figure S1C) indicated the presence of local extremes instead of abundance over the whole cell area. These local extremes were pronounced for the groups of MOI 0.5 and Cyt-D. These results agree with our observation of cortical rings as a dense accumulation of actin filaments at the cell periphery. Due to the denser presence of actin and possible formation of aggregates, the fluorescence intensity in these locations was higher (Figure S2).

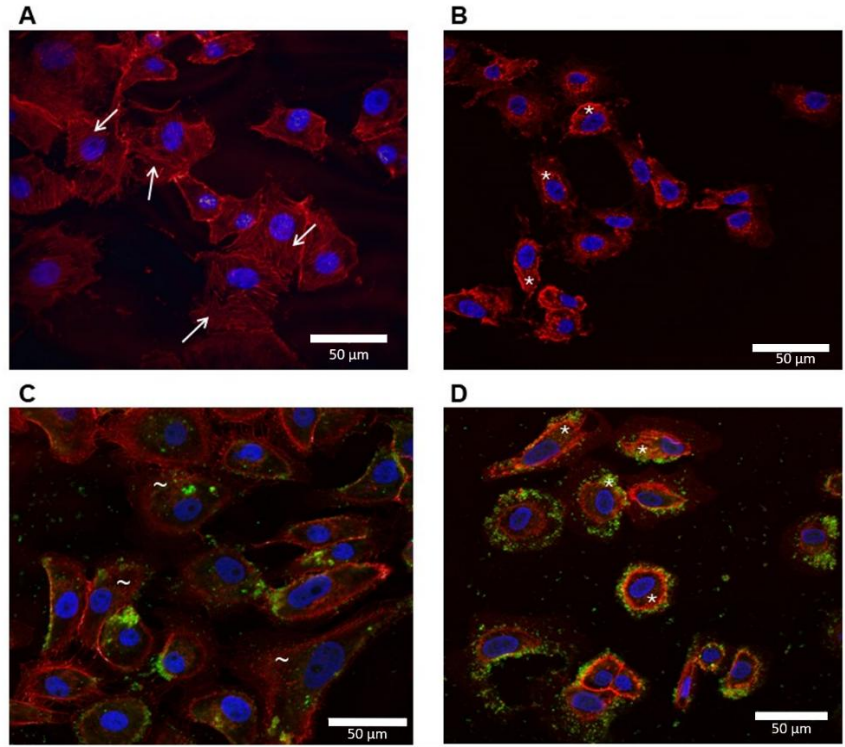

**Figure 3.** Fluorescence images of Vero CCL-81 cells stained with Phalloidin, showing actin (red); DAPI, showing cell nuclei (blue) and measles virus F-antigen (green) under different conditions. Non-infected cells (A) were used as negative control showing an intact cytoskeletal structure with thick F-actin bundles which are indicated by the white arrows, spanning the cell body. Cytochalasin D treated cells (B) were used as positive control displaying disrupted F-actin organization, loss of stress fibers and shrunken cell bodies (*). Cells infected with MVV at MOI 0.1 (C) show partial cortical accumulation of F-actin with residual internal stress fibers and focal adhesions marked with "~" to show affected cells, indicating cytoskeletal remodeling. Cells infected with MVV at MOI 0.5 (D) show actin fibers predominantly redistributed to the cell periphery, forming cortical actin rings with a notable absence of internal stress fibers and shrinked cell bodies (*).

## 4. Discussion

In this work, we investigated MVV virus infection induced changes in Vero cells from a biomechanical perspective. The results showed that infection is associated to a reduction in cellular stiffness, originating from a change in the structure and organization of actin filaments, as quantified by AFM measurements and complementary confocal microscopy analysis. These findings are consistent with previous reports (2).

Mechanical measurements revealed that cells infected with a higher MVV dose (MOI 0.5) experienced greater softening reduction in their Young's modulus compared to cells infected at MOI 0.1, 24 hours post infection (Figure 1A). The results were compared to those obtained from cells treated with the actin depolymerizer Cyt-D, as well as to uninfected cells. Cyt-D treated cells presented mechanical properties like those observed at MOI 0.5. In contrast, cells infected at a lower MOI showed stiffness values like uninfected cells. The significant decrease in Young's modulus observed for cells exposed to MOI 0.5 reflected a cytoskeletal disruption as fluorescence microscopy images illustrated in Figure 3. Further mechanical description should consider the viscoelastic nature of the cells.

Thus, viscoelastic parameters are shown in Figure 2. The viscosities extracted from the Maxwell model ($\eta_1$ and $\eta_2$) decreased systematically in infected and cytochalasin D treated cells compared with mock controls (Figure 2A). The observed trend suggests viscous dissipation and faster mechanical relaxation. Accordingly, Maxwell elements support that infection induces a perinuclear shift in viscoelastic response rather than affecting a single parameter. Additionally, rather than treating $\beta$ as a first indicator of cellular fluidity, the present study considers it a descriptor of the time-dependent viscoelastic response of the probed cellular region. Consistently, $\beta$ increased in MOI 0.5 and Cyt-D treated cells compared to mock controls, indicating faster stress relaxation dynamics. However, no statistically significant difference was observed between mock and MOI 0.1 conditions for $\beta$ (Figure 2B).

In parallel, morphological and fluorescence-based studies revealed infection-associated redistribution of actin related signals. Cells infected at higher MOI exhibited a significant reduction in cell area accompanied by decreased mean fluorescence intensity and increased maximum fluorescence intensity, consistent with the formation of localized high-intensity regions (Figure S1). Together, these findings suggest that infection promotes reorganization of actin into discrete subcellular domains (24, 25). The high variability in cell area observed in the non-infected and MOI 0.1 groups reflects greater morphological heterogeneity associated with a healthy cellular state (Figure S1A). These cells remained elongated, motile, and proliferative, contributing to a wider range of cell sizes. Additionally, the MOI 0.1 condition could represent a mixed population of infected as well as non-infected cells. In contrast, infection at the higher MOI (0.5) resulted in uniformly reduced cell areas with lower variability, consistent with infection-induced cell rounding and a loss of morphological diversity, a phenotype comparable to that observed following Cyt-D treatment. Next, we ascribed the higher MFI in mock-infected cells (Figure S1B) to the fact that F-actin stayed intact in those the untreated cells and the mode of the fact that Phalloidin staining, binds to whole actin filaments instead of monomers, which are present more frequently in the treated cells.

Structural changes of the actin filaments revealed dose-dependent alterations in filament length and width (Figure S2). Cells infected at MOI 0.5 showed reduced filament length and increased heterogeneity in filament width, similar to the changes observed after Cyt-D treatment. These quantitative analysis of actin filaments further supports the AFM results, indicating substantial reorganization of actin architecture during infection.

## 5. Conclusions

The findings of this work show that Vero cells exposed to MVV undergo virus dose-dependent softening and changes in viscoelastic properties within 24 hours of infection.

The mechanical behavior observed in cells exposed to MVV are similar to the results obtained by disrupting actin filaments with Cyt-D. Combined atomic force microscopy and confocal imaging revealed that these changes coincide with actin redistribution from intracellular stress fibers to cortical rings, indicating substantial cytoskeletal remodeling. Together, these results provide a quantitative description of the coupling between cytoskeletal organization and local viscoelastic properties in measles virus–infected cells.

Further studies incorporating time-resolved measurements, infection efficiency metrics, and mechanistic perturbations will be required to achieve a more complete description of the infection mechanism.

## Acknowledgements

This research was funded by the COMET-Funding Program, managed by the Austrian Research Promotion Agency FFG and by the Austrian Science Fund (FWF project W1224 - Doctoral Program Biomolecular Technology of Proteins). The authors thank Lukas Schrangl for reading the manuscript, and Amsatou Andorfer Sarr and Valerie Wagner for technical assistance.

# Appendix

## Contents

# S1. Materials and methods

## S.1.1 Cell culture and infection with measles vaccine virus (MVV)

Vero CCL-81 cells were cultured in Dulbecco's modified Eagle's medium (DMEM, high glucose, GlutaMAX™; Gibco, Thermo Fisher Scientific Cat. No. 10566016, USA) supplemented with 10% fetal bovine serum (Gibco™, Thermo Fisher Scientific, Cat. No. A5256701, USA) and 1% penicillin-streptomycin (Gibco™, Thermo Fisher Scientific, Cat. No. 15140122, USA) at 36.5°C in 5% $CO_2$.

For infection experiments, cells were plated to ~50% confluence and infected with MVV at a desired multiplicity of infection (MOI). The details of the production of MVV and the analysis of infectious virus particles by TCID50 were described elsewhere (1-3). The MOIs used for infection in this study were 0.1 and 0.5 and were generated by using a purified MVV virus stock with an initial concentration of $10^5$ TCID50/mL.

After virus exposure for 24 hours at 36.5°C, infected cells were used for mechanical measurements. Non-infected cells were used as controls. In some experiments, a subset of these cells was treated with cytochalasin D (C8273, Sigma-Aldrich, Merck; 5 µM for 30 minutes prior to measurement) to serve as a positive control for cytoskeletal disruption. Cytochalasin D is a mycotoxin that binds to the barbed ends of actin filaments, preventing their polymerisation and causing disruption to the actin network.

## S.1.2 Atomic force microscopy (AFM)

For AFM measurements, cells were seeded at a density of $5 \times 10^5$ cells in 35 mm untreated glass-bottom Petri dishes (Ibidi) and incubated overnight before treatment. Cells were passaged at approximately 80% confluency, and for infection experiments cells were exposed to MVV at multiplicities of infection (MOIs) of 0.1 or 0.5. After 24 h of virus exposure at 36.5 °C, infected Vero cells were subjected to mechanical measurements, with non-infected cells serving as negative controls and cells treated with cytochalasin D (5 µM, 30 min) as a positive control for cytoskeletal disruption. During measurements, physiological conditions were maintained using Leibovitz's L-15 medium in a temperature-controlled stage set to 37 °C. The AFM probe was positioned over the central nuclear region using transmission light microscopy; each cell was measured in triplicate, with at least 30 cells per biological replicate and a total of four biological replicates. A sample was kept in the sample stage for a maximum of 2 h, during measuring time no visible changes were appreciated on the cell morphology.

Tipless triangular cantilevers (MLCT-O10, Bruker) with a nominal spring constant of 0.1 N/m and resonance frequency of 38 KHz were converted into colloidal probes by fixing a 10 µm diameter silica bead (microParticles GmbH) to the tip, following the protocol outlined by Weber et al. (4). Prior to measurements, probes were cleaned with EtOH, dried with $N_2$ and cleaned with 30 min of UV/$O_3$, then calibrated by acquiring force-distance-curves on a stiff substrate (glass) and using the thermal noise method provided by JPK software.

AFM force-distance measurements were performed using a JPK NanoWizard III atomic force microscope (AFM; Bruker, Germany) integrated with a with a CellHesion module on top of a

Zeiss inverted optical microscope (Axio Observer Z1, Zeiss, Jena, Germany), the stage was a temperature-controlled PetriDishHeater (Bruker).

Force-distance measurements (Scheme S1A) start with the colloidal probe approaching the cells at a rate of 1 μm/s until it reaches the designated setpoint (1nN), then the cantilever retracts at a loading rate of 1 μm/s from the sample. Some cells may adhere to the probe after the measurement leaving adhesive interactions. The Z range of approach and retract was 10 μm and the data was recorded at a sample rate of 5000 Hz. Data obtained from Force-Distance measurements (Scheme S1B) were used to extract the young's Modulus by using the Hertz model with substrate thickness correction.

Stress relaxation measurements (Scheme S1C) start with the cantilever approaching the cell at 1 μm/s until it makes contact with the cell surface and reaches a setpoint of 1 nN, then the position is held at constant height for 5 s (to observe viscoelastic deformation), and finally the cantilever was retracted at 1 μm/s, adhesive interactions between the tip and the sample were detected as downward force deflections before complete detachment. The Z range of approach and retract was 3 μm and the data was recorded at a sample rate of 5000 Hz. Data obtained (Scheme S1D) from the pause segment at constant height was fitted to a 5-element Maxwell model and a phenomenological power law model.

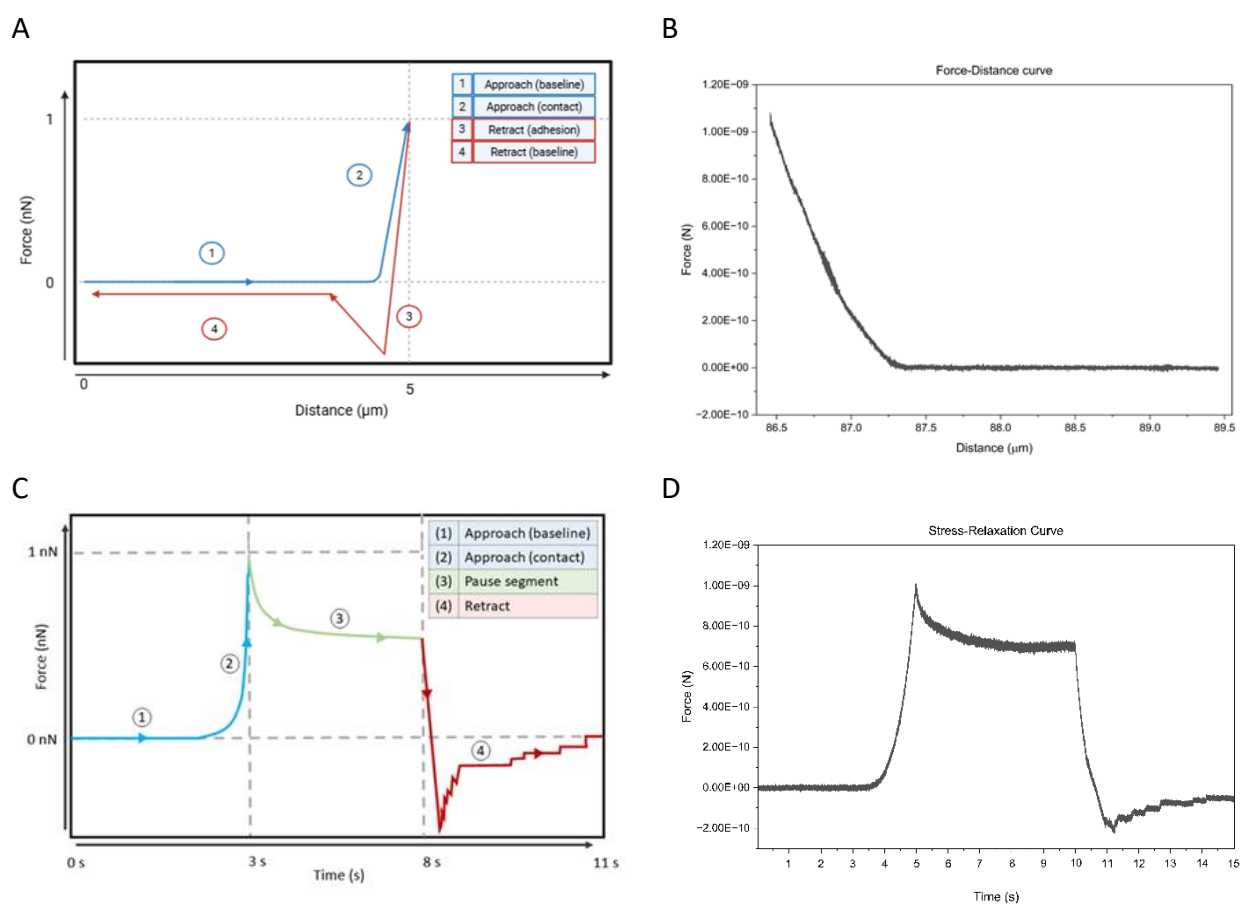

**Scheme S1**. (A) AFM force–distance curve in contact mode. The measurement comprises four distinct segments: (1) an initial approach phase without contact, establishing the baseline; (2) continued approach until the cantilever contacts the cell surface and reaches the predefined setpoint force; (3) a retraction phase, during which adhesive interactions between the probe and cell surface become apparent; and (4) a final retraction segment where the probe fully detaches, returning to a new

baseline. (B) Representative approach and contact segment of a force-distance curve measured on an infected cell. (C) AFM stress–relaxation force curve consisting of four distinct segments: (1) an initial approach without contact, establishing the baseline; (2) continued approach until the cantilever contacts the cell surface and reaches the setpoint force; (3) a dwell phase, during which the force gradually relaxes due to viscoelastic deformation of the sample; and (4) cantilever retraction, where adhesive interactions are observed prior to complete detachment. (D) Representative stress-relaxation force curve measured on an infected cell.

### S.1.3. Data processing and analysis

The force-distance and force-time curves were processed using a custom software toolkit in R (5). The batch process included baseline-correction, contact point determination and zero-point slope determination.

### S.1.3.1 Evaluation of the indentation

The indentation depth ($\delta$) into the cell was calculated as the difference between the cantilever deflection ($\delta_c$) and piezo extension ($Z_p$) as shown in the following equation:

$$[\delta = Z_p - \delta_c = Z_p - \frac{F}{k_c}] \tag{S1}$$

where F is the measured force and $k_c$ the spring constant of the cantilever.

### S.1.3.2 Determination of the apparent Young's modulus

In a first approximation to evaluate the modulus of Young of the cells, the first 500 nm of the indentation segment of the force-distance curve were fitted considering the influence of the sample thickness by using the following model (6):

$$F = \frac{16}{9} E \sqrt{R \delta^3} [1 + 1.113\chi + 1.283X^2 + 0.769\chi^3 + 0.0975\chi^4] \tag{S2}$$

With E being the apparent Young's Modulus, R the radius of the particle, and $\chi$ being $\sqrt{\frac{R\delta}{h}}$ where h (5 µm) represents the cell thickness. This model considers if substrate effects are relevant based on cell thickness.

### S.1.3.3 Viscoelastic analysis

Cells possess complex and heterogeneous internal architectures composed of cytoskeletal networks, membrane compartments, and organelles. Consequently, they exhibit viscoelastic behaviour rather than purely elastic or purely viscous responses (7-9). Stress-relaxation experiments were employed to quantify the cellular response to a constant applied strain and to characterize time-dependent mechanical processes associated with cytoskeletal rearrangement and intracellular dissipation. To probe this behaviour, a constant indentation depth ($\delta_0$) was applied, and the indentation was approximated by a Heaviside step function, allowing the measured force response to be directly related to the time-dependent elastic modulus.

For a spherical indenter, the force response can be written as

$$F(t) = \frac{C}{1-\nu} E(t) \tag{S3}$$

where E(t) is the relaxation modulus, ν is the Poisson ratio, and C is a geometry-dependent constant given by

$$C = \frac{4}{3}\sqrt{R}\delta_0^{3/2} \tag{S4}$$

Here, C represents the purely geometric contribution of the spherical particle-cell contact. Importantly, C is independent of the material properties and remains constant during the relaxation phase, such that all time dependence in F(t) arises from the viscoelastic response of the sample. The viscoelastic response was modelled using a generalized Maxwell framework consisting of an equilibrium elastic element in parallel with two Maxwell branches. The relaxation modulus is expressed as

$$E(t) = E_\infty + \sum_{i=1}^{2} E_i e^{\left(-\frac{t}{\tau_i}\right)} \tag{S5}$$

where $E_\infty$ is the equilibrium modulus, $E_i$ are the moduli associated with each Maxwell element, and $\tau_i$ are the corresponding relaxation times. Substituting this expression into the force equation yields:

$$F(t) = \frac{C}{1-\nu}\left(E_\infty + E_1 e^{\left(-\frac{t}{\tau_1}\right)} + E_2 e^{\left(-\frac{t}{\tau_2}\right)}\right) \tag{S6}$$

This formulation explicitly separates geometry (through C) from material behaviour (through $E_\infty, E_i, \tau_i$), enabling a direct physical interpretation of the fitted parameters. The viscosity of the maxwell branches (η) is calculated as

$$\eta_i = E_i \tau_i \tag{S7}$$

The two relaxation times capture fast and slow dissipative processes, commonly associated with cortical and cytoplasmic mechanical rearrangements, respectively (10, 11).

In addition to the discrete viscoelastic models, a phenomenological power-law rheological model was employed to describe the stress-relaxation behaviour of cells without assuming a finite number of characteristic relaxation times. This approach has been applied to living cells (12). Thus, under stress-relaxation conditions at constant indentation, the force response was modelled as

$$F(t) = \frac{8}{3}\sqrt{R\delta^3}\, E_0 \left(\frac{t}{t_0}\right)^{-\beta} \tag{S8}$$

where $t_0=1s$ is a normalization time, $E_0$ is the effective relaxation modulus at $t=t_0$, and $\beta$ is the power-law exponent.

The parameter $E_0$ represents an effective stiffness of the cell at one second after deformation, providing a standardized measure for comparison across experimental conditions. The exponent $\beta$ characterizes the breadth of the viscoelastic relaxation spectrum: values approaching $\beta=0$ correspond to predominantly elastic, solid-like behaviour, whereas values approaching $\beta=1$ indicate increasingly fluid-like behaviour dominated by viscous dissipation. Unlike generalized Maxwell models, which assume a finite number of discrete relaxation times, power-law models capture a continuous distribution of relaxation processes that is characteristic of the complex, hierarchical structure of living cells. Soft biological materials such as the cytoskeleton exhibit viscoelastic behavior governed by a broad spectrum of relaxation time scales arising from interactions among polymer networks, transient crosslinks, and active remodeling processes, which are not well described by a small set of exponential relaxation modes (13, 14).

### S.1.3.4 Statistical analysis

Statistical analyses were performed using R (version 4.x). The Shapiro–Wilk test was used to test for normality, which indicated that cell stiffness values were not normally distributed. Therefore, non-parametric tests were applied. Differences between two groups were evaluated using the two-sided Wilcoxon rank-sum test, with $p < 0.05$ being considered significant. For multiple comparisons (e.g. comparing all treatment groups with the control group), a Bonferroni correction ($p < 0.01$) was applied.

### S.1.4 Confocal microscopy

To visualize cytoskeletal organization, Vero cells were seeded at a density of 4-5 x 10⁴ cells on sterile chamber slides (µ-Slide 4-well, ibidi) and infected with MVV at MOI 0.1 or 0.5 under the same conditions as above. At 24 h post-infection, cells were gently washed with phosphate-buffered saline (PBS) and fixed in 4% paraformaldehyde (10 min, room temperature). After three PBS washes, cells were permeabilized with 0.1% Triton X-100 (15 min) and blocked with 2% bovine serum albumin (BSA) in PBS (overnight at 4°C). The actin cytoskeleton was stained with phalloidin (Alexa Fluor 555 conjugate, 1:400) for 30 min in the dark. In parallel, to confirm infection, some samples were immuno-stained with a primary antibody against measles virus F-antigen (by Thermo Fisher Scientific, polyclonal rabbit, 1:100 in 1% BSA, 3 h) followed by a co-staining of Alexa Fluor 633-labelled secondary antibody (1:1000, 30 min). Cell nuclei were counterstained with DAPI (1 µg/mL, 5 min). Confocal images were captured on a Leica TCS SP8 LIGHTNING confocal microscope (Leica Microsystems, Wetzlar, Germany) with a 40x oil immersion objective under constant exposure settings for comparison between conditions. Quantitative analysis of confocal images was performed with ImageJ software.

# S2. Calculated elastic and viscoelastic parameters

**Table S1.** Calculated elastic and viscoelastic parameters. N stands for sample size, SEM for standard error of the mean, STD for standard deviation and $R^2$ for the coefficient of determination with respective standard deviation.

Young's Modulus (E)

| Condition | N   | Mean (Pa) | SEM (Pa) | STD (Pa) | Median (Pa) | $R^2$         |
|-----------|-----|-----------|----------|----------|-------------|---------------|
| mock      | 126 | 701.7     | 19.6     | 219.7    | 687.7       | 0.994 ± 0.007 |
| MOI 0.1   | 199 | 643.3     | 21.8     | 308.1    | 610.4       | 0.992 ± 0.009 |
| MOI 0.5   | 294 | 421.2     | 16.4     | 281.1    | 363.8       | 0.985 ± 0.027 |
| Cyt-D     | 264 | 321.4     | 17.1     | 278.5    | 241.3       | 0.992 ± 0.013 |

Indentation (δ)

| Condition | N   | Mean (µm) | SEM (µm) | STD (µm) | Median (µm) | $R^2$        |
|-----------|-----|-----------|----------|----------|-------------|--------------|
| mock      | 126 | 0.26      | 0.02     | 0.20     | 0.26        | 0.943 ±0.023 |
| MOI 0.1   | 199 | 0.34      | 0.02     | 0.23     | 0.33        | 0.955 ±0.027 |
| MOI 0.5   | 294 | 0.42      | 0.02     | 0.34     | 0.36        | 0.966 ±0.024 |
| Cyt-D     | 264 | 0.36      | 0.02     | 0.28     | 0.27        | 0.974 ±0.029 |

Equilibrium modulus ($E_\infty$)

| Condition | N   | Mean (Pa) | SEM (Pa) | STD (Pa) | Median (Pa) | $R^2$        |
|-----------|-----|-----------|----------|----------|-------------|--------------|
| mock      | 74  | 218.49    | 10.31    | 88.65    | 213.93      | 0.943 ±0.023 |
| MOI 0.1   | 63  | 159.30    | 14.06    | 111.62   | 125.44      | 0.955 ±0.027 |
| MOI 0.5   | 143 | 106.49    | 6.77     | 81.00    | 84.41       | 0.966 ±0.024 |
| Cyt-D     | 163 | 79.62     | 6.56     | 83.72    | 37.65       | 0.974 ±0.029 |

Elastic modulus 1 ($E_1$)

| Condition | N   | Mean (Pa) | SEM (Pa) | STD (Pa) | Median (Pa) | $R^2$        |
|-----------|-----|-----------|----------|----------|-------------|--------------|
| mock      | 74  | 53.21     | 2.98     | 21.71    | 51.97       | 0.943 ±0.023 |
| MOI 0.1   | 63  | 36.45     | 4.47     | 27.00    | 28.55       | 0.955 ±0.027 |
| MOI 0.5   | 143 | 24.51     | 3.92     | 19.40    | 19.31       | 0.966 ±0.024 |
| Cyt-D     | 163 | 21.31     | 4.01     | 18.50    | 12.17       | 0.974 ±0.029 |

Elastic modulus 2 ($E_2$)

| Condition | N | Mean (Pa) | SEM (Pa) | STD (Pa) | Median (Pa) | $R^2$ |
|---|---|---|---|---|---|---|
| mock | 74 | 103.19 | 7.37 | 74.90 | 88.26 | 0.943 ±0.023 |
| MOI 0.1 | 63 | 71.36 | 6.65 | 56.17 | 54.04 | 0.955 ±0.027 |
| MOI 0.5 | 143 | 52.11 | 7.97 | 57.51 | 34.02 | 0.966 ±0.024 |
| Cyt-D | 163 | 34.11 | 5.36 | 31.30 | 18.24 | 0.974 ±0.029 |

Relaxation time 1 ($\tau_1$)

| Condition | N | Mean (s) | SEM (s) | STD (s) | Median (s) | $R^2$ |
|---|---|---|---|---|---|---|
| mock | 74 | 0.205 | 0.020 | 0.175 | 0.176 | 0.943 ±0.023 |
| MOI 0.1 | 63 | 0.177 | 0.018 | 0.141 | 0.145 | 0.955 ±0.027 |
| MOI 0.5 | 142 | 0.283 | 0.022 | 0.267 | 0.210 | 0.966 ±0.024 |
| Cyt-D | 163 | 0.201 | 0.009 | 0.116 | 0.181 | 0.974 ±0.029 |

Relaxation time 2 ($\tau_2$)

| Condition | N | Mean (s) | SEM (s) | STD (s) | Median (s) | $R^2$ |
|---|---|---|---|---|---|---|
| mock | 74 | 4.10 | 0.44 | 3.74 | 2.66 | 0.943 ±0.023 |
| MOI 0.1 | 63 | 2.80 | 0.31 | 2.50 | 2.21 | 0.955 ±0.027 |
| MOI 0.5 | 143 | 3.18 | 0.20 | 2.44 | 2.48 | 0.966 ±0.024 |
| Cyt-D | 163 | 2.53 | 0.15 | 1.93 | 1.98 | 0.974 ±0.029 |

Viscosity 1 ($\eta_1$)

| Condition | N | Mean (Pa s) | SEM (Pa s) | STD (Pa s) | Median (Pa s) | $R^2$ |
|---|---|---|---|---|---|---|
| mock | 74 | 10.79 | 0.94 | 8.08 | 9.33 | 0.943 ±0.023 |
| MOI 0.1 | 63 | 6.34 | 0.83 | 6.59 | 5.08 | 0.955 ±0.027 |
| MOI 0.5 | 141 | 6.77 | 0.75 | 8.93 | 4.04 | 0.966 ±0.024 |
| Cyt-D | 162 | 4.28 | 0.38 | 4.83 | 2.45 | 0.974 ±0.029 |

Viscosity 2 ($\eta_2$)

| Condition | N | Mean (Pa s) | SEM (Pa s) | STD (Pa s) | Median (Pa s) | $R^2$ |
|---|---|---|---|---|---|---|
| mock | 74 | 501.67 | 94.88 | 816.17 | 198.90 | 0.943 ±0.023 |
| MOI 0.1 | 62 | 204.66 | 32.96 | 259.52 | 118.77 | 0.955 ±0.027 |
| MOI 0.5 | 142 | 183.52 | 29.57 | 352.31 | 87.60 | 0.966 ±0.024 |
| Cyt-D | 163 | 116.77 | 19.48 | 248.71 | 37.21 | 0.974 ±0.029 |

Effective relaxation modulus ($E_0$)

| Condition | N | Mean (Pa) | SEM (Pa) | STD (Pa) | Median (Pa) | $R^2$ |
|---|---|---|---|---|---|---|
| mock | 100 | 288.78 | 9.73 | 97.29 | 285.89 | 0.884 ± 0.026 |
| MOI 0.1 | 116 | 237.46 | 13.66 | 147.16 | 202.04 | 0.875 ± 0.040 |
| MOI 0.5 | 225 | 147.71 | 6.90 | 103.49 | 114.20 | 0.883 ± 0.039 |
| Cyt-D | 197 | 129.07 | 8.31 | 116.59 | 59.14 | 0.905 ± 0.040 |

Power law exponent (β)

| Condition | N | Mean | SEM | STD | Median | $R^2$ |
|---|---|---|---|---|---|---|
| mock | 100 | 0.072 | 0.002 | 0.019 | 0.072 | 0.884 ± 0.026 |
| MOI 0.1 | 116 | 0.081 | 0.002 | 0.023 | 0.078 | 0.875 ± 0.040 |
| MOI 0.5 | 225 | 0.084 | 0.002 | 0.027 | 0.083 | 0.883 ± 0.039 |
| Cyt-D | 197 | 0.096 | 0.002 | 0.025 | 0.097 | 0.905 ± 0.040 |

# S3. Quantitative analysis of confocal images

A

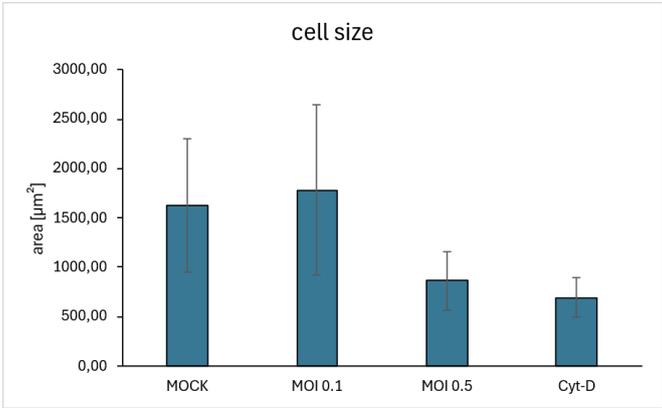

B

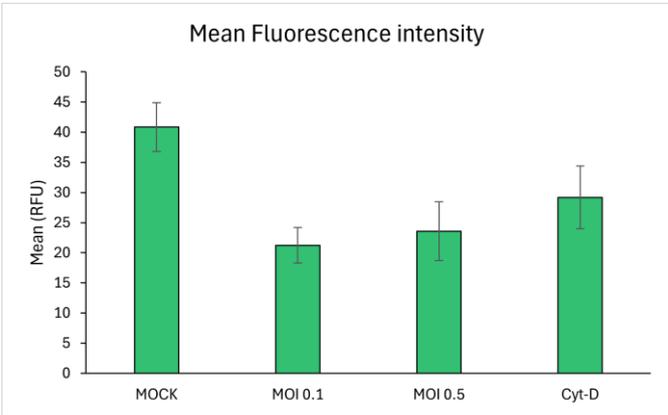

C

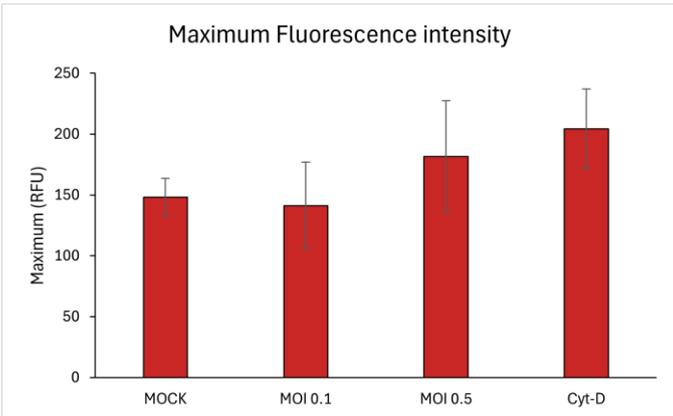

**Figure S1.** Quantitative data of confocal images of the investigated groups – mock, MOI 0.1, MOI 0.5 and Cyt-D. 10 cells per image were used to measure cell size (A), mean fluorescence intensity (B) and max fluorescence intensity (C) in relative fluorescence units, RFU.

## S4. Filament sensor images

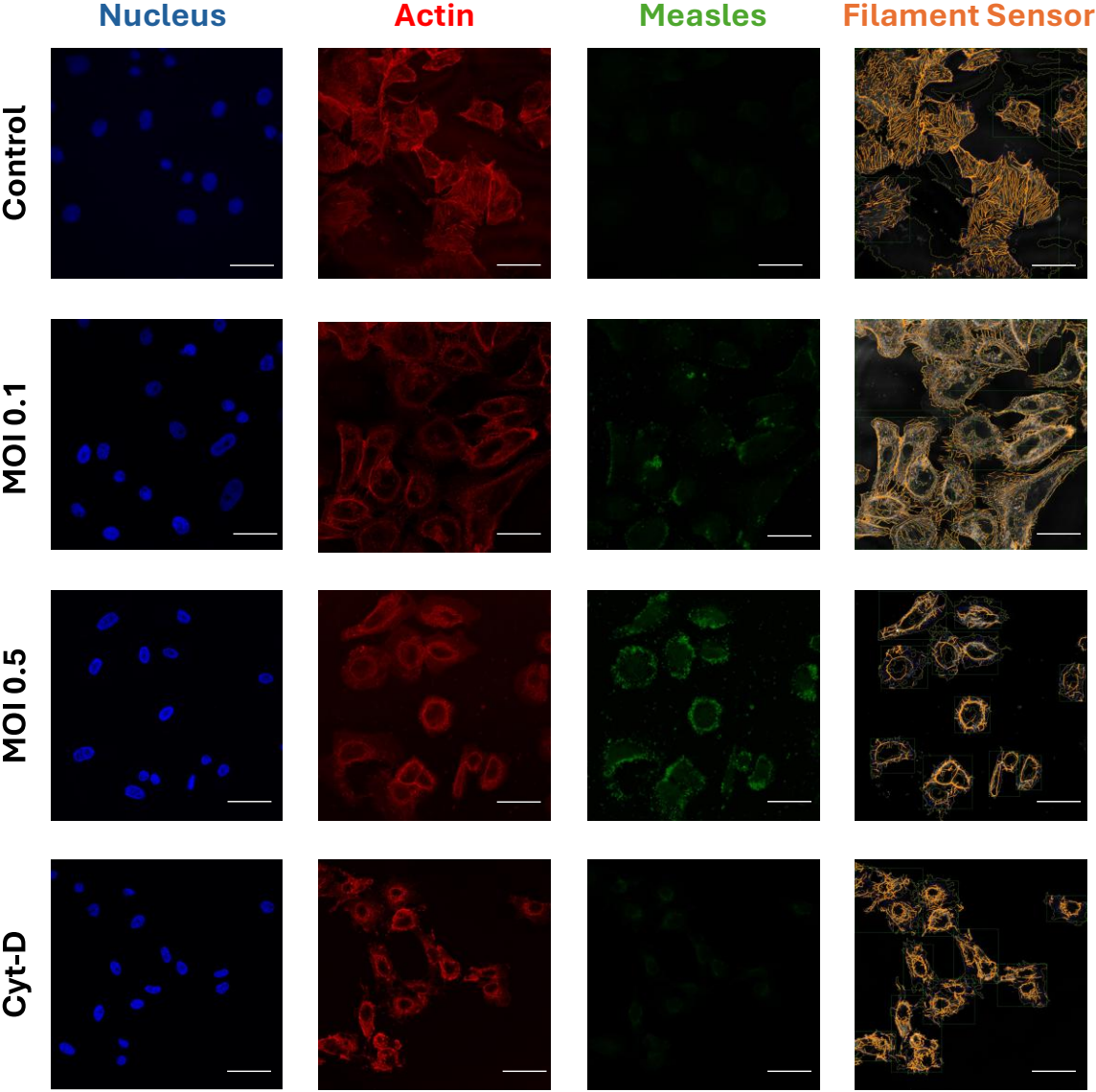

**Figure S2.** Representative immunofluorescence Confocal Laser Scaning Microscope (CLSM) images of each of the treated conditions displaying the nuclei in blue, the actin cytoskeleton in red and the MVV in green. In orange we can observe the results from introducing the image through Filament sensor (15) and obtaining the filament distribution. In the images there is a clear difference in actin distribution as well as MVV presence between different treatments. White scale bar represents 50 µm.